\documentclass[12pt]{article}

\usepackage{scicite}
\usepackage{times}
\usepackage{amsmath,amssymb,graphicx,fullpage,color,mathtools,amsthm,xcolor}
\usepackage{caption,subcaption}
\usepackage{algorithm,algorithmic}
\usepackage[normalem]{ulem}

\newtheorem*{theorem*}{Theorem}
\newtheorem*{corollary*}{Corollary}

\usepackage{microtype}
\usepackage{hyperref,color}

\definecolor{webgreen}{rgb}{0,.35,0}
\definecolor{webbrown}{rgb}{.6,0,0}
\definecolor{RoyalBlue}{rgb}{0,0,0.9}
\definecolor{purp}{rgb}{0.6,0.05,0.8}
\definecolor{ora}{rgb}{0.7,0.35,0.02}

\hypersetup{
   colorlinks=true, linktocpage=true, pdfstartview=FitV,
   breaklinks=true, pdfpagemode=UseNone, pageanchor=true, pdfpagemode=UseOutlines,
   plainpages=false, bookmarksnumbered, bookmarksopen=true, bookmarksopenlevel=1,
   hypertexnames=true, pdfhighlight=/O,
   urlcolor=webbrown, linkcolor=RoyalBlue, citecolor=webgreen,
   pdfauthor={Levi H. Dudte, Gary P. T. Choi, L. Mahadevan},
   pdfsubject={An additive algorithm for origami design}
}

\newenvironment{sciabstract}{%
\begin{quote} \bf}
{\end{quote}}

\newcounter{lastnote}

\title{An additive algorithm for origami design}

\author{Levi H. Dudte$^{1}$, Gary P. T. Choi$^{1,2}$, L. Mahadevan$^{1,3\ast}$\\
\\
\footnotesize{$^{1}$John A. Paulson School of Engineering and Applied Sciences, Harvard University}\\
\footnotesize{$^{2}$Department of Mathematics, Massachusetts Institute of Technology}\\
\footnotesize{$^{3}$Departments of Physics, and Organismic and Evolutionary Biology, Harvard University}\\
\footnotesize{$^\ast$To whom correspondence should be addressed; E-mail: lmahadev@g.harvard.edu}
}

\date{}

\begin{document} 


\baselineskip24pt


\maketitle 

\begin{sciabstract}
Inspired by the allure of additive fabrication, we pose the problem of origami design from a new perspective: how can we grow a folded surface in three dimensions from a seed so that it is guaranteed to be isometric to the plane? We solve this problem in two steps: by first identifying the geometric conditions for the compatible completion of two separate folds into a single developable four-fold vertex, and then showing how this foundation allows us to grow a geometrically compatible front at the boundary of a given folded seed. This yields a complete marching, or additive, algorithm for the inverse design of the complete space of developable quad origami patterns that can be folded from flat sheets. We illustrate the flexibility of our approach by growing ordered, disordered, straight and curved folded origami and fitting surfaces of given curvature with folded approximants. Overall, our simple shift in perspective from a global search to a local rule has the potential to transform origami-based meta-structure design.
\end{sciabstract}

\begin{figure*}[t!]
\centering
\includegraphics[width=0.95\textwidth]{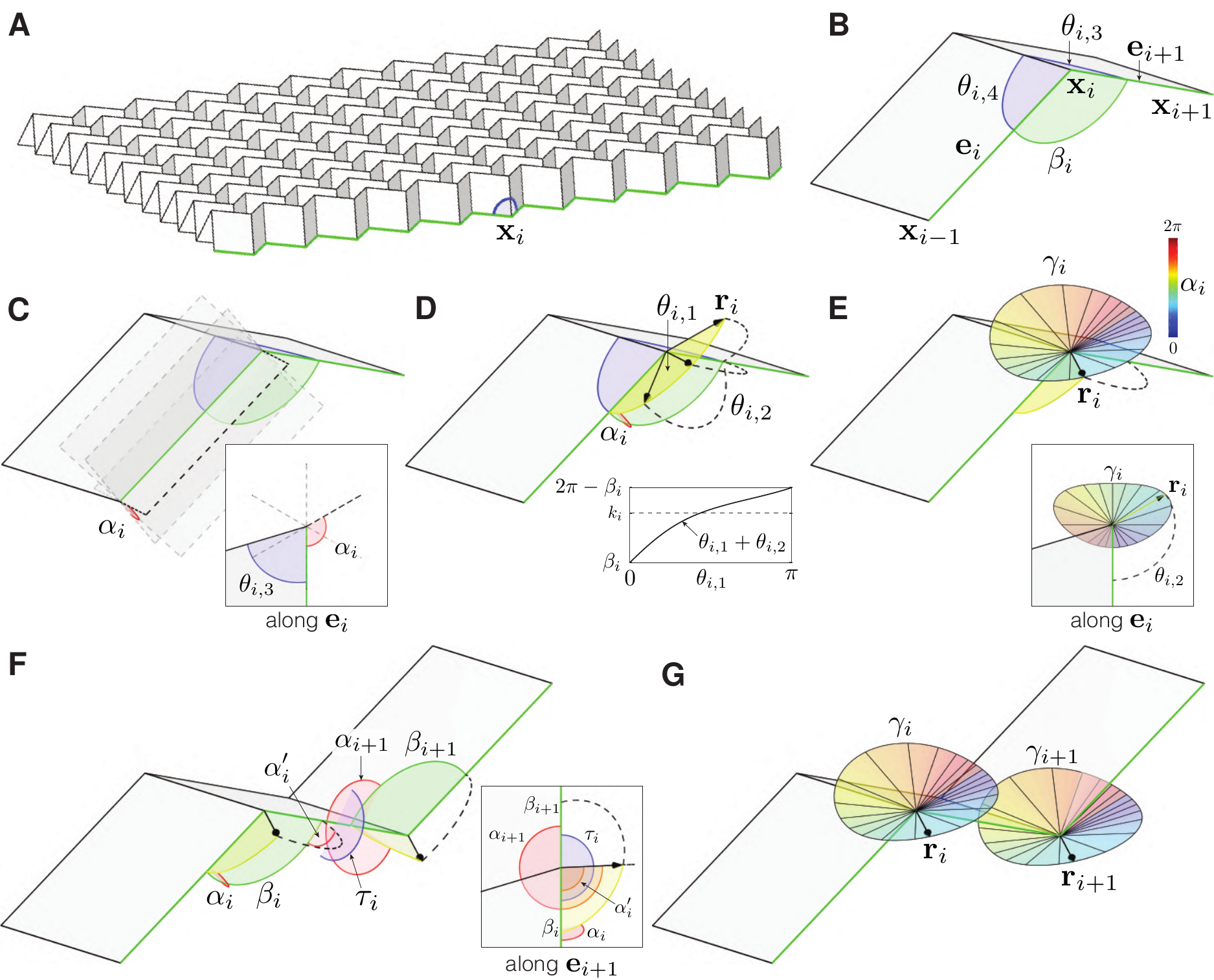}
\caption[]{
\textbf{Construction of Quad Origami.}
\textbf{(A)} A quad origami surface, the Miura-ori pattern, with boundary vertex $\mathbf{x}_i$ along a growth front (green).
\textbf{(B)} Focusing on $\mathbf{x}_i$, shown from a different vantage point than that of (A), its adjacent growth front vertices $\mathbf{x}_{i-1}$ and $\mathbf{x}_{i+1}$. The two design angles along the boundary $\theta_{i,3}$ and $\theta_{i,4}$ are shown in blue and the angle in space $\beta_{i}$ between the two growth front edges $\mathbf{e}_i$ and $\mathbf{e}_{i+1}$ in green.
\textbf{(C)} The plane of action for new design angle $\theta_{i,1}$ is determined by a flap angle $\alpha_i$ (red), which sweeps from the $\beta_i$ face clockwise about $\mathbf{e}_i$.
\textbf{(D)} As $\theta_{i,1}$ (yellow) sweeps through its plane of action, it determines possible growth directions $\mathbf{r}_i$ and $\theta_{i,2}$ (dashed), the angle between $\mathbf{r}_i$ and $\mathbf{e}_{i+1}$. These must satisfy $\theta_{i,1}+\theta_{i,2}=k_i$ (inset) to create a developable vertex $\mathbf{x}_i$. 
\textbf{(E)} This constraint gives an ellipse $\gamma_i$ of spherical arcs $\theta_{i,1}$ and $\theta_{i,2}$ which forms a closed loop around the line containing $\mathbf{e}_i$. The value of $\theta_{i,1}$ which satisfies the constraint is given by the unique intersection of its plane of action and $\gamma_i$, so $\alpha_i$ parameterizes $\gamma_i$.
\textbf{(F)} The secondary flap angle $\alpha_{i}'$ at $\mathbf{x}_i$ sweeps from the $\beta_{i}$ face clockwise about $\mathbf{e}_{i+1}$ and is determined by $\alpha_i$. The flap angle $\alpha_{i+1}$ at $\mathbf{x}_{i+1}$ sweeps from the $\beta_{i+1}$ face clockwise about $\mathbf{e}_{i+1}$ to the same plane as measured by $\alpha_i'$. 
\textbf{(G)} Two adjacent growth directions $\mathbf{r}_i$ and $\mathbf{r}_{i+1}$ must be coplanar, so $\mathbf{r}_{i+1}$ is determined by the intersection of this plane and $\gamma_{i+1}$, thus $\alpha_{i+1}$ parameterizes $\gamma_{i+1}$.
}
\label{fig:construction}
\end{figure*}

Folding patterns arise in nature in systems including insect wings, leaves and guts~\cite{kobayashi1998the,bowden1998spontaneous,mahadevan2005self,shyer2013} and have a long history in decorative, ceremonial and pedagogical traditions of origami around the world. More recently, they have begun to draw the attention of mathematicians fascinated by the patterns and limits of folding~\cite{huffman76primer,liu2019invariant,lebee2018fitting,demaine_orourke,lang2011origami} and engineers and scientists inspired by their technological promise~\cite{dias2012geometric,wei2013geometric,liu2018topological,santangelo2017extreme,filipov2015origami,chen2015origami,demaine2017origamizer,callens2018from,he2019approximating}. 
\\
\\
The simplest origami is a single vertex with four folds, a kind of hydrogen atom of folding with exactly one internal degree of freedom. Patterns comprised of four-coordinated vertices and quadrilateral faces are called quad origami, which may have isolated folded configurations isometric to the plane, if they can be folded at all. The mechanical response of structures and materials derived from quad origami is governed in large part by geometric frustration encountered during folding. Using these patterns to program rigid-foldable and flat-foldable, floppy or multistable systems then requires consideration of additional symmetries~\cite{pinson2017self} and folds~\cite{chen2019percolation}, making quad origami a promising platform for meta-structures at any scale from the nanoscopic to the architectural. This has attracted significant scientific interest to the problem of their design, but the challenge of either finding quad patterns that actually fold or, inversely, surfaces that unfold has limited freeform solutions.
\\
\\
Previous quad origami design studies have tended to focus on tessellations with periodic geometries and specific mountain/valley (MV) assignments assumed {\it a priori}, the well-known Miura-ori pattern~\cite{miura1980method} being the canonical example, and have generally employed either direct geometric methods to parameterize simple design variations~\cite{mitani2009design,gattas2013miura,gattas2014miura,zhou2015design,sareh2015design,wang2016folding,song2017trapezoidal,hu2019design,feng2020helical} or optimization algorithms to generalize known folding typologies~\cite{tachi2009generalization,dudte2016programming,kilian2008curved,jiang2019curve}. The former typically provide a comprehensive understanding of a restricted space of designs sharing strong qualitative similarities, i.e. those exhibiting particular symmetries, and involve constructions that are inevitably case-specific. The latter typically require encoding nonlinear developability constraints in a non-convex, multidimensional optimization framework and use a well-known periodic folding pattern as an empirical departure point. These computational methods are generic, but they suffer from two problems: the difficulty of finding a good guess to ensure convergence to a desired local solution and the lack of scalability to large problems. Thus while many current strategies have been used to expound on a wide variety of quad origami patterns, the general problem of quad origami design has admitted only piecemeal solutions and the science has for the most part followed the art form.
\\
\\
Inspired by the simple edge extrusion operation from computational design and additive fabrication, one can ask the following inverse problem: how we can extend the boundary of a folded quad origami surface outward such that the new surface remains developable, i.e. isometric to the plane and thus capable of being fabricated from flat sheets. In the case of origami this implies that a folding process can transform the pattern through intermediate configurations to a second global energy minimum, the designed, folded surface. Although this typically implies geometric frustration in intermediate stages, recently, several marching algorithms have been developed to design quad origami which, in addition to developing to the plane, can deploy rigidly, i.e. with no geometric frustration. This allows for deployment from a flat to folded to flat-folded configurations with one degree of freedom, a sub-class of developable quad origami known as rigid- and flat-foldable~\cite{lang2018rigidly,dieleman2020jigsaw,feng2020rffo}, such as a ``jigsaw'' method to design rigid-foldable quad origami~\cite{dieleman2020jigsaw}, a combinatorial strategy borrowed from artistic modular origami design~\cite{fuse1990unit}, wherein geometrically compatible folded units are selected from a predetermined set of discrete modules to augment the boundary of a folded bulk model. 
\\
\\
Here we deviate fundamentally from these previous approaches by providing a unified framework that identifies the complete continuous family of compatible, folded strips that can be extruded directly from the boundary of a folded model. This lays the foundation for a simple geometric additive algorithm that allows us to explore the entire space of developable quad origami designs, not limited to just rigid- and/or flat-foldable designs. We begin by exploring the flexibility in angles and lengths associated with fusing two pairs of folded faces at a common boundary which yields the geometric compatibility conditions for designing a four-coordinated single vertex origami. We then apply the single-vertex construction to determine the space of compatible quad origami strips at the boundary of an existing folded model. Critically, we establish that the new interior edge directions and design angles along the growth front form a one-dimensional family parameterized by the choice of a single face orientation in space along the growth front. The result is an additive geometric algorithm for the evolution of folded fronts around a prescribed seed into a folded surface, establishing the means to characterize the full design space of generic quad origami surfaces. This constructive algorithm~\cite{dudte2017thesis} is enabled by the following:
\begin{theorem*}
The space of new interior edge directions along the entire growth front in a quad origami is one-dimensional, i.e. uniquely determined by a single angle.
\end{theorem*}
{\bf Proof.}   Our proof primarily consists of three parts: single vertex construction, construction of adjacent vertices and the growth of the full growth, with details given in SI Section S1.
\\
\\
{\it 1. Single vertex construction}: Suppose we are given a vertex along the growth front with position vector $\mathbf{x}_i$ (Fig.~\ref{fig:construction}A), with the two adjacent growth front vertices denoted by $\mathbf{x}_{i-1}, \mathbf{x}_{i+1}$. We denote the two boundary design angles incident to $\mathbf{x}_i$ in the existing surface by $\theta_{i,3}$ and $\theta_{i,4}$ and the angle in space at $\mathbf{x}_i$ along the growth front denoted by $\beta_i = \angle \{-\mathbf{e}_i, \mathbf{e}_{i+1}\} \in (0,\pi)$ (Fig.~\ref{fig:construction}B), where $\mathbf{e}_{i} = \mathbf{x}_i - \mathbf{x}_{i-1}$ and $\mathbf{e}_{i+1} = \mathbf{x}_{i+1} - \mathbf{x}_i$. To obtain a new edge direction vector $\mathbf{r}_i$ that gives the direction of an interior edge $[\mathbf{x}_i, \mathbf{x}_i']$ in the augmented quad origami surface, let $\alpha_i \in [0,2\pi)$ be the left-hand oriented flap angle about $\mathbf{e}_i$ from the $\beta_i$ plane to the plane of the new quad containing $\mathbf{r}_i$ and $\mathbf{e}_i$ (Fig.~\ref{fig:construction}C). We note that the single vertex origami at $\mathbf{x}_i$ satisfies the local angle sum developability condition
\begin{align}
\sum_{j=1}^4 \theta_{i,j} = 2\pi,
\label{eq:angle_sum}
\end{align}
where $\theta_{i,1} = \angle \{-\mathbf{e}_i, \mathbf{r}_{i}\} \in (0,\pi)$ and $\theta_{i,2} = \angle \{\mathbf{e}_{i+1}, \mathbf{r}_{i}\} \in (0,\pi)$ are two new design angles implied by $\mathbf{r}_i$ (Fig.~\ref{fig:construction}D). Furthermore, $\theta_{i,1}$, $\theta_{i,2}$ and $\beta_i$ form a spherical triangle with $\alpha_i$ an interior spherical angle opposite $\theta_{i,2}$, so that the spherical law of cosines gives
\begin{align}
\cos{\theta_{i,2}} = \cos{\theta_{i,1}}\cos{\beta_i}+\sin{\theta_{i,1}}\sin{\beta_i}\cos{\alpha_i}.
\label{eq:spherical_triangle}
\end{align}
Solving Eq.~\eqref{eq:angle_sum} and Eq.~\eqref{eq:spherical_triangle} for $\theta_{i,1}$, $\theta_{i,2}$ yields 
\begin{align}
{\theta_{i,1}} &= \tan^{-1} \frac{\cos{k_i}-\cos{\beta_i}}{\sin{\beta_i}\cos{\alpha_i} - \sin{k_i}}, ~~\theta_{i,1} \neq \pi/2 \label{eq:theta1}\\
\theta_{i,2} &= k_i - \theta_{i,1},
\label{eq:theta2}
\end{align}
where $k_i = 2\pi - \theta_{i,3} - \theta_{i,4}$, the amount of angular material required to satisfy developability. If $\theta_{i,1} = \pi/2$, we have $\cos \theta_{i,2} = \sin \beta_i \cos \alpha_i$, which yields a unique solution if $\beta_i \neq 0, \pi$ and $\beta_i < k_i < 2\pi-\beta_i$ (see SI Section S1 for details). We thus see that the solutions $\theta_{i,1}, \theta_{i,2}$ to Eq.~\eqref{eq:angle_sum} and Eq.~\eqref{eq:spherical_triangle} exist and are unique for any given $\theta_{i,3}$, $\theta_{i,4}$ and $\beta_i$ (angles intrinsic to the existing origami) and $\alpha_i$ (the flap angle), modulo a finite number of singular configurations. The new transverse edge direction $\mathbf{r}_i$ can then be obtained using $\theta_{i,1}$ and $\theta_{i,2}$ (Fig.~\ref{fig:construction}E). The key geometric intuition and an alternative proof of existence and uniqueness of single vertex solutions is to observe that $k_i$ defines an ellipse $\gamma_i$ of spherical arcs $\theta_{i,1}, \theta_{i,2}$ that satisfy Eq.~\eqref{eq:angle_sum} with foci given by $-\mathbf{e}_i$ and $\mathbf{e}_{i+1}$. For any flap angle $\alpha_i$, the sum $\theta_{i,1} + \theta_{i,2} = \beta_i$ when $\theta_{i,1}=0$ and $\theta_{i,1} + \theta_{i,2} = 2\pi-\beta_i$ when $\theta_{i,1}=\pi$ and the sum $\theta_{i,1} + \theta_{i,2}$ is positive monotonic on the interval $\theta_{i,1}\in[0,\pi]$, generically. Moreover, the spherical triangle inequality bounds $\beta_i<\theta_{i,3} + \theta_{i,4}< 2\pi-\beta_i$, generically, so no matter what flap angle is chosen or inherited from a neighboring vertex a unique solution to Eq.~\eqref{eq:angle_sum} and Eq.~\eqref{eq:spherical_triangle} must exist and the flap angle $\alpha_i$ parameterizes the ellipse $\gamma_i$. See SI Section~S1 and Movie~S1 for further details and discussion.
\\
{\it 2. Construction of adjacent vertices:} We now show that the new edge directions $\mathbf{r}_{i+1}$, $\mathbf{r}_{i-1}$ at $\mathbf{x}_{i+1}$, $\mathbf{x}_{i-1}$ are also uniquely determined by the single flap angle $\alpha_i$.
Without loss of generality, consider obtaining $\mathbf{r}_{i+1}$ given $\mathbf{r}_{i}$ (Fig.~\ref{fig:construction}F). Denote $\alpha_i'$ as the left-hand oriented angle about $\mathbf{e}_{i+1}$ from the $\beta_i$ plane to the plane of the new quad containing $\theta_{i,2}$. Referring again to the spherical triangle formed by $\theta_{i,1}$, $\theta_{i,2}$ and $\beta_i$, the spherical laws of sines and cosines give
\begin{align}
\sin{\alpha_i'} &= \frac{\sin\theta_{i,1}(\alpha_i)}{\sin\theta_{i,2}(\alpha_i)}\sin\alpha_{i}, \label{eq:sin_alpha_prime} \\
\cos{\alpha_i'} &= \frac{\cos{\theta_{i,1}(\alpha_i)} - \cos{\theta_{i,2}(\alpha_i)}\cos{\beta_i}}{\sin{\theta_{i,2}(\alpha_i)}\sin{\beta_i}}, \label{eq:cos_alpha_prime}
\end{align}
yielding a unique solution $\alpha_i' \in [0,2\pi)$. As $\theta_{i,1}$ and $\theta_{i,2}$ are functions of $\alpha_i$, $\alpha_i'$ is also a function of $\alpha_i$. Observe that $\alpha_i'$ and $\alpha_{i+1}$ are measured about a common axis and are thus related by a shift of the left-hand oriented angle $\tau_i$ from the $\beta_i$ face to the $\beta_{i+1}$ face. This gives the flap angle transfer function $g_i:[0,2\pi) \to [0,2\pi)$:
\begin{equation}
\alpha_{i+1} = g_i(\alpha_i) = \text{mod}(\alpha_i'(\alpha_i) - \tau_i, 2\pi)
\end{equation}
as measured left-hand oriented about $\mathbf{e}_{i+1}$ starting at the $\beta_i$ plane. It is easy to see that $g$ is bijective, hence $\mathbf{r}_{i+1}$ is uniquely determined by $\alpha_i$ and both $\gamma_i$ and $\gamma_{i+1}$ are parameterized by $\alpha_i$ (Fig.~\ref{fig:construction}G). A similar argument applies for $\mathbf{r}_{i-1}$. For geometric intuition, observe that there are bijections between points on $\gamma_i$ and the half-planes about $\mathbf{e}_i$ and $\mathbf{e}_{i+1}$.
\\
{\it 3. Growth of the entire front}: Finally, to establish bijection between the flap angles $\alpha_i$ and $\alpha_j$ at arbitrary $i,j$ where $i < j$, we consider the following composition $f_{i,j}$ of the transfer functions $g$:
\begin{align}
\alpha_{j} = f_{i,j}(\alpha_i) = g_j(g_{j-1}(g_{j-2}(\cdots g_i(\alpha_i)))).
\end{align}
Since each transfer function is bijective, their composition is also bijective. Therefore, all new interior edge directions along the entire growth front are parameterized by a single flap angle $\alpha_i$. \hfill $\blacksquare$ 

\begin{figure*}[t]
\centering
\includegraphics[width=\textwidth]{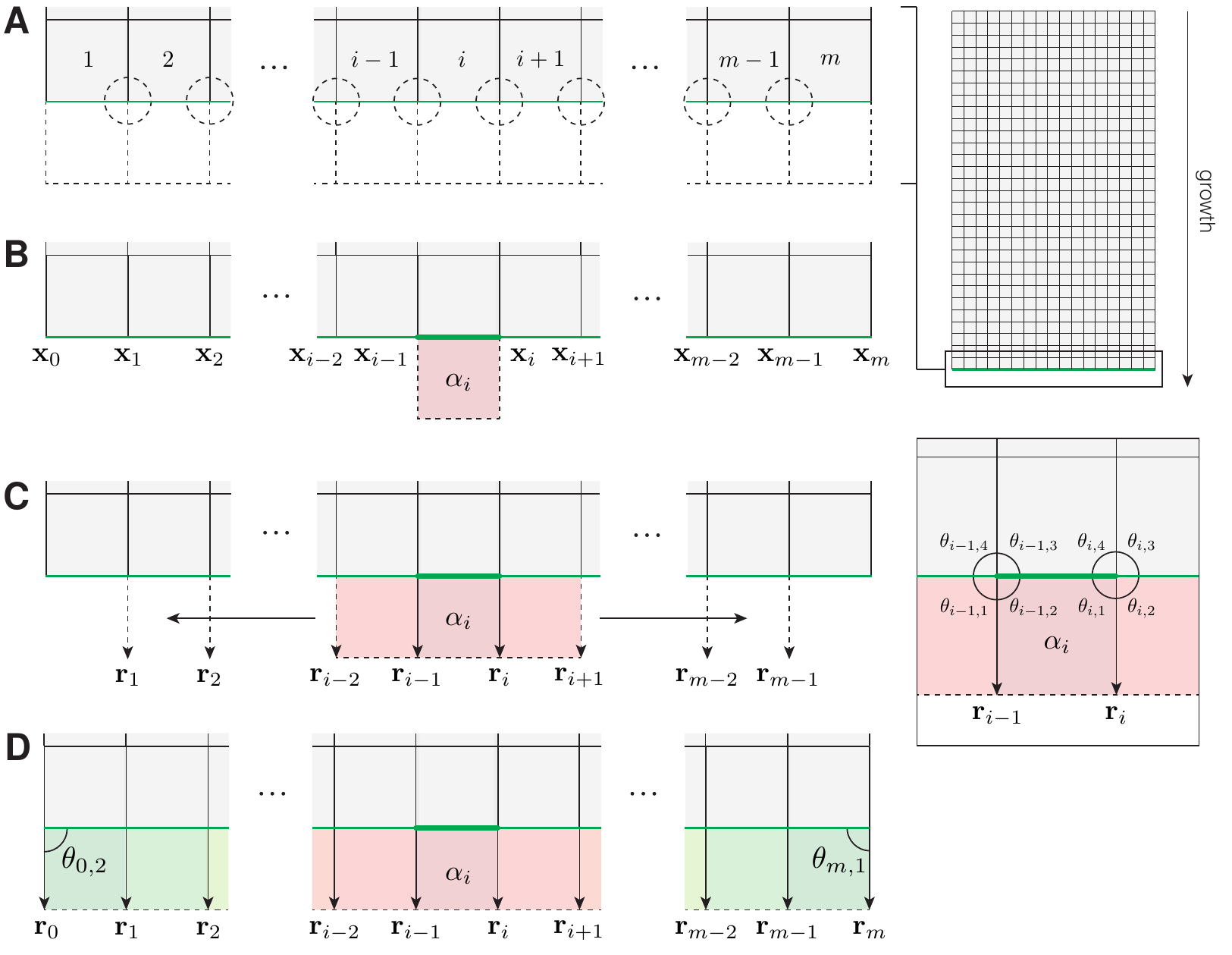}
\caption[]{\textbf{Additive Algorithm.}
\textbf{(A)} To grow an existing folded quad origami model at a boundary having $m$ quads, $m+1$ new vertices must be placed in space subject to $m$ planarity constraints (dashed squares) and $m-1$ angle sum constraints (dashed circles), for a total of $3(m+1) - (2m-1)$ = $m+4$ DOFs, generically.
\textbf{(B)} The additive strip construction begins by choosing the plane associated with any one of the quad faces in the new strip (1 DOF).
\textbf{(C)}~Consecutive single vertex systems propagate this flap angle choice down the remainder of the strip, determining uniquely the orientations in space of all quad faces in the new strip.
\textbf{(D)}~Edge directions at the endpoints of the strip can be chosen freely in their respective planes (2 DOFs) and all transverse edges in the new strip can be assigned lengths ($m+1$ DOFs) for a total $m+4$ DOFs.}
\label{fig:algorithm}
\end{figure*}

\begin{figure*}[t]
\centering
\includegraphics[width=\textwidth]{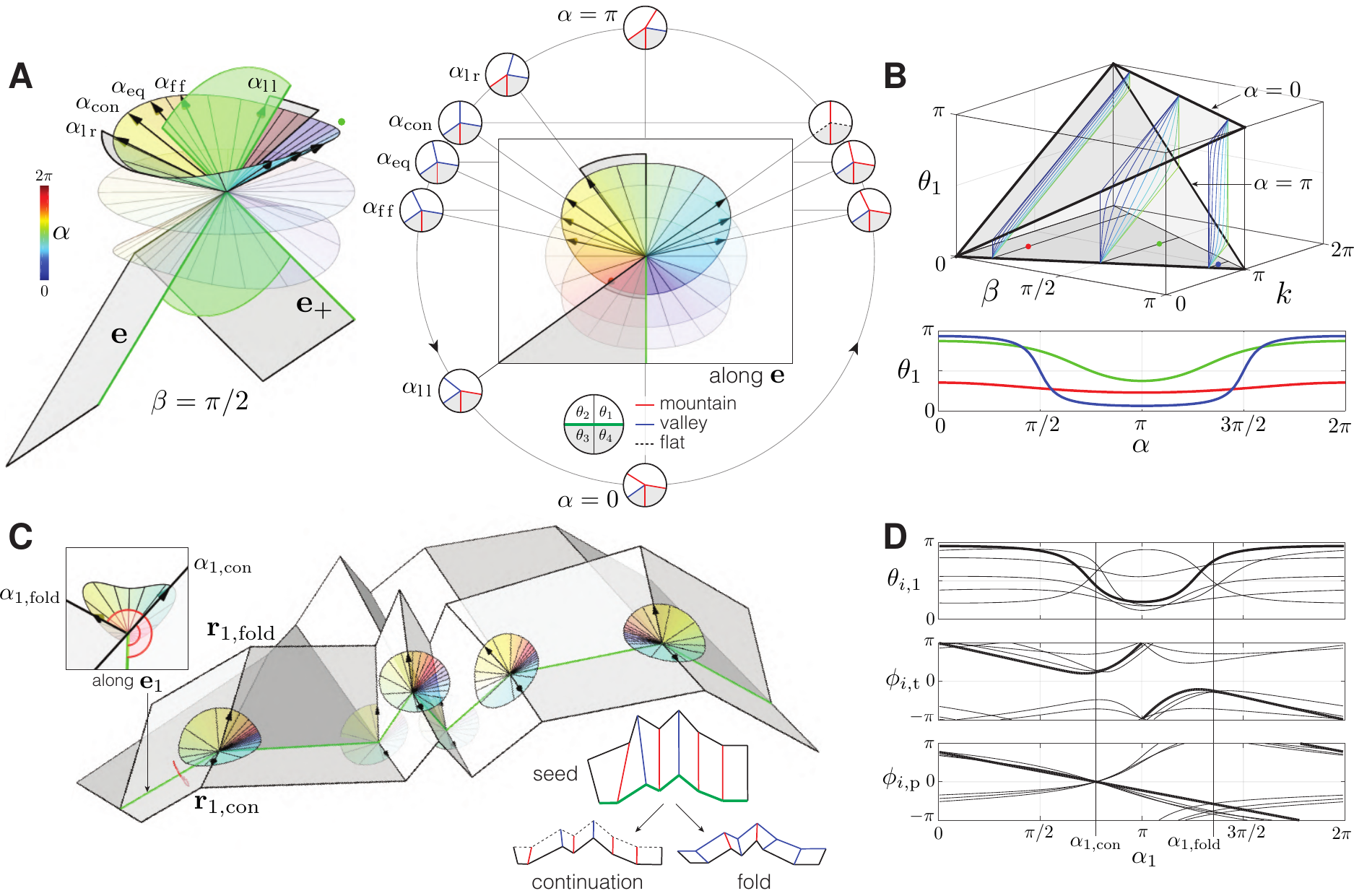}
\caption[]{\textbf{Vertex and Strip Design.} \textbf{(A)} A pair of folded quads and its single vertex growth front with $\beta=\pi/2$ and the spherical ellipse given by $k=5\pi/4$ are shown, along with two other faint ellipses given by $k=\pi,3\pi/4$ that would be given by different existing design angles than those shown. Special vertex growth directions and self-intersection intervals are recovered by identifying their flap angles, three of which ($\alpha_{\text{con}}, \alpha_{\text{eq}}, \alpha_{\text{ff}}$) have two solutions given by reflection over the $\beta$ plane. \textbf{(B)} Valid region for $\beta$ and $k$ at a single vertex (top). Typical generic growth front vertices (red, green and blue points) fall in the interior of this region. The first design angle $\theta_1$ is bounded above by $(k + \beta)/2$ at $\alpha = 0$ and below by $(k - \beta)/2$ at $\alpha = \pi$ and surfaces of constant $\alpha$ are shown in the interstice. Sweeping $\alpha \in [0,2\pi)$ produces possible values for $\theta_1$ (bottom) symmetric about $\alpha=\pi$ for the three colored points identified above. \textbf{(C)}~A folded quad strip with two compatible growth directions (continuation, where no new fold is created, and a folded configuration) selected from the one-dimensional space of compatible strip designs parameterized by the orientation in space of the first new face. \textbf{(D)}~Half of the new interior design angles $\theta_{i,1}$ in the new strip (top), fold angles transverse to the growth front $\phi_{i,t}$ (middle) and parallel to the growth front $\phi_{i,p}$ (bottom) are shown as functions of flap angle $\alpha_1$. Characteristic single vertex curves associated with $\mathbf{x}_1$ are bolded while curves associated with other vertices (light black) differ from characteristic single vertex curve shapes by non-locality.}
\label{fig:strip}
\end{figure*}

\begin{figure*}[t]
\centering
\includegraphics[width=\textwidth]{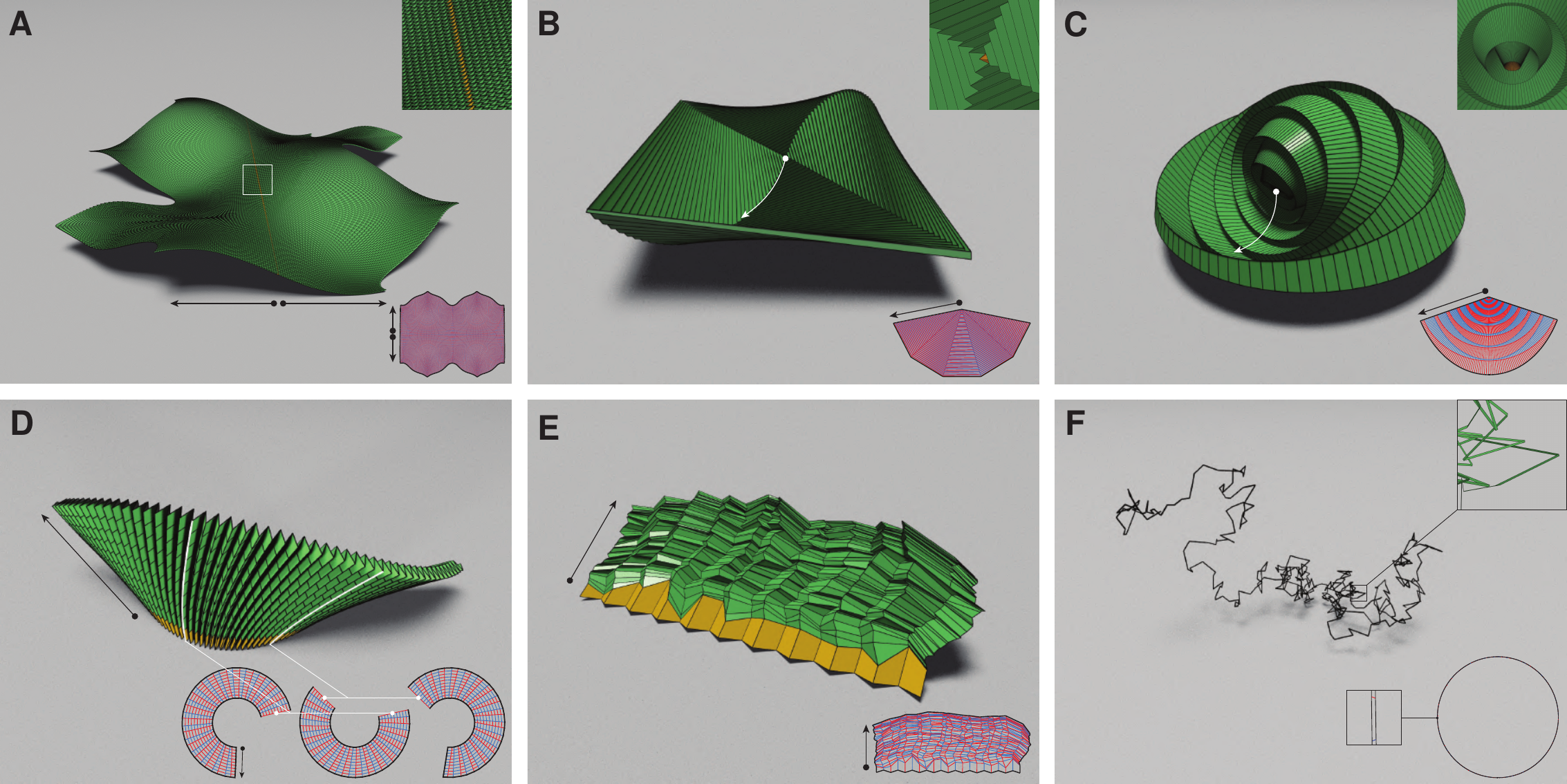} %F4
\caption[]{\textbf{Additive Design of Straight, Curved, Ordered and Disordered Origami.}
\textbf{(A)} A generalized Miura-ori tessellation fit to a target surface with mixed Gaussian curvature generated. Lower and upper bounding surfaces are displaced normally from the target surface and a seed strip of quads is initialized in between the two with one growth front on each surface. New strips are attached on either side of the seed by reflecting the growth front back and forth between bounding surfaces in their interstice.
\textbf{(B)} A low-resolution conical seed with four-fold symmetry grows by reflecting between the interstices of two rotating upper and lower boundary planes. New strips form closed loops with overlapping endpoint faces.
\textbf{(C)} A high-resolution conical seed grows by attaching progressively tilted cone rings to reproduce a curved-fold model. New strips form closed loops with overlapping endpoint faces.
\textbf{(D)} A curved fold model grows from a seed created using the corollary to attach a strip of quads to a corrugated parabola.
\textbf{(E)} A self-avoiding walk away from a Miura-ori seed strip with noise added to the boundary growth front produces a crumpled sheet. New strips are added by sampling flap angles within bounds that prevent local self-intersection.
\textbf{(F)} A Brownian ribbon whose development approximates a circular annulus is created using the corollary.
The seeds are highlighted in yellow, and the arrows indicate the growth direction. The fold pattern for each model is shown at the bottom right of each image. See SI Figs.~S8 and S9 for higher-resolution versions of the fold patterns.
}
\label{fig:surface}
\end{figure*}

\begin{corollary*} Given a generic curve $\mathcal{C}$ discretized by $m+1$ vertices $\mathbf{x}_i\in\mathbb{R}^3, i = 0,\dotsc,m$ and $m$ edges $\mathbf{e}_i=\mathbf{x}_i-\mathbf{x}_{i-1}, i= 1,\dotsc,m$, with angles $\beta_i=\angle \{-\mathbf{e}_i, \mathbf{e}_{i+1}\} \in (0,\pi), i= 1,\dotsc,m-1$, the space of planar patterns that fold to $\mathcal{C}$ is $m$-dimensional.
\end{corollary*}
{\bf Proof.} Consider assigning $k_i \in (\beta_i,2\pi-\beta_i), i= 1,\dotsc,m-1$ to the interior vertices of $\mathcal{C}$. In the above origami proof, $\mathcal{C}$ is a growth front and $k_i$ are given by the existing origami surface. For a discrete curve, $k_i$ can be chosen freely to determine a one-dimensional set of fold directions $\mathbf{r}_i\in \mathbb{R}^3, i= 1,\dotsc,m-1$ that give a development of $\mathcal{C}$ to the plane. \hfill $\blacksquare$
\\
\\
This proof suggests immediately an efficient geometric algorithm for designing generic quad origami surfaces. For an existing regular quad origami (seed) with a growth front designated by a strip of $m$ boundary quads (Fig.~\ref{fig:algorithm}A), we note that a new strip of $m$ quads has $3(m+1)$ DOFs in $\mathbb{R}^3$ subject to $m$ planarity and $m-1$ design angle constraints. If we add a new strip to a boundary with $m$ quads, we have a total of $m+4$ DOFs to determine the geometry of the new strip: $1$ flap angle to determine the interior design angles and edge directions, $2$ boundary design angles at the endpoints of the strip, and $m+1$ edge lengths. So while our main theorem establishes the design space of generic quad origami, the following geometric algorithm allows us to explore this landscape additively, satisfying developability constraints by construction along the way. A new compatible strip of $m$ quads is designed by the following steps.
\begin{enumerate}
\item Start from any $i \in \{1, 2, \dots, m-1\}$ and choose the flap angle $\alpha_{i}$ associated with the growth front edge $\mathbf{e}_i$ ($1$~DOF) (Fig.~\ref{fig:algorithm}B).
\item Propagate the $\alpha_i$ choice along the growth front from $\mathbf{x}_i$ to $\mathbf{x}_{1}$ and $\mathbf{x}_{m-1}$ by iteratively solving for $\theta_{i,1}, \theta_{i,2}$, rotating $\mathbf{r}_i$, calculating $\alpha_{i-1}, \alpha_{i+1}$, and moving on to the next vertex (Fig.~\ref{fig:algorithm}C).
\item Choose boundary design angles $\theta_{0,2}, \theta_{m,1} \in (0,\pi)$ and rotate $\mathbf{r}_0$ and $\mathbf{r}_m$ into position ($2$~DOFs) (Fig.~\ref{fig:algorithm}D).
\item Calculate the new edge length bounds and choose $l_j$ for all $j$ ($m+1$ DOFs). Bounds are given by the observation that the new outward facing edges in each new quad cannot intersect each other, which occurs when the two interior angles of a new quad sum to less than $\pi$ (see SI Section~S3).
\item Calculate the new vertex positions given $\mathbf{r}_j$ and $l_j$ for all $j$.
\item Repeat the above steps at any boundary front to grow more new strips.
\end{enumerate}
The algorithm also applies to discrete curves not associated with an existing folded surface via the corollary. In this case, we can design the shape of the development of the curve by choosing $k$ values in Step 2, rather than calculating them from an existing surface.
\\
\\
Having established generic connections from single vertices to quad strips to origami surfaces, we now analyze the flap angle parameterization at each scale of this hierarchy in more detail. The design space of the growth front of a pair of folded faces, a proto-single vertex origami, is described fully by the pair of scalars $\beta$, the angle in space formed by the growth front, and $k$, the shape parameter of $\gamma$, i.e. the amount of angular material required to satisfy developability (Fig.~\ref{fig:strip}A). A generic single vertex origami can be constructed in the interior of the triangular region $0\le\beta\le\pi$ and $\beta\le k\le 2\pi - \beta$, with singular configurations at the boundaries given by equality (Fig.~\ref{fig:strip}B). Sweeping $\alpha$ from $0$ to $2\pi$ parameterizes the ellipse such that $\theta_{1}(\alpha=0)=(k+\beta)/2$ and $\theta_{1}(\alpha=\pi)=(k-\beta)/2$ and new edge directions $\mathbf{r}(\alpha)$ tend to cluster in space around growth front directions where $d\theta_{1}/d\alpha$ has smaller magnitude. Special single vertex origami~\cite{waitukaitis2016origami} are recovered by identifying their flap angles (Fig.~\ref{fig:strip}A, see SI for formulae). Three of these vertex types are given by rearranging Eq.~\eqref{eq:theta1} and plugging in a desired value for the new design angle: the continuation solution $\alpha_{\text{con}}$ where the new pair of quads can be attached without creating a new fold, the flat-foldable solution $\alpha_{\text{ff}}$ where the vertex can be fully folded such that all faces are coplanar and $\alpha_{\text{eq}}$ which creates equal new design angles. These each admit two solutions related by reflection over the plane of the growth front, recovering a duality noted in~\cite{huffman76primer}. Two more special vertices are identified by $\alpha_{\text{ll}}$ and $\alpha_{\text{lr}}$, which produce locked configurations with the left pair of faces $(\theta_1, \theta_4)$ and the right pair $(\theta_2, \theta_3)$ folded to coplanarity, respectively. A vertex is trivially locked with coplanar $(\theta_1, \theta_2)$ faces when $\alpha = 0, \pi$. The continuation flap angle that does not create a new fold along the growth front and the locked-left flap angle are related by $\alpha_{\text{con}} = \text{mod}(\alpha_{\text{ll}}+\pi,2\pi)$. We also note that self-intersection will occur for flap angles in the intervals between the $\beta$ plane and the nearest non-trivial locked flap angle. Notably absent from our construction are fold angles, which can be recovered at growth front edge $\mathbf{e}_i$ by $\phi_{i,\text{p}}=\alpha_{i,\text{ll}}-\alpha_i-\pi$.
\\
\\
Moving up in scale, we explore the relationship between flap angle and strip design. The special single vertex solutions cannot necessarily be enforced at all locations along a generic surface growth front as the space of growth directions is one-dimensional. To design a set of flat-foldable growth front folds, for example, requires additional symmetries. The exception to this is $\alpha_{i, \text{con}}$, the single flap angle value that gives the trivial growth direction for the entire front. In Fig.~\ref{fig:strip}C, we illustrate a generic folded quad strip and two of its compatible strips, the trivial continuation solution and a non-trivial folded solution. New design angles $\theta_{i,1}$, fold angles transverse to the growth front $\phi_{i,\text{t}}$ and parallel to the growth front $\phi_{i,\text{p}}$ in the new strip are shown as functions of flap angle $\alpha_1$ in Fig.~\ref{fig:strip}D. Fold angles parallel to the growth front $\phi_{i,\text{p}}$ are simultaneously zero at $\alpha_{1, \text{con}}$ and non-zero otherwise, while fold angles transverse to the growth front $\phi_{i,\text{t}}$ are generically never zero. See SI Section~S2 and Movies~S2--S3 for more details.
\\
\\
To show the capability of our additive approach, we now deploy it in inverse design frameworks to construct ordered and disordered quad origami typologies with straight and curved folds. In contrast with previous work~\cite{dudte2016programming}, our additive approach does not require the solution of a large multi-dimensional optimization problem for the entire structure. Instead it only requires choosing from the available DOFs for each strip, which map the full space of compatible designs, and hence is more computationally feasible and geometrically complete. These choices are application-specific, and can be random, interactive or based on some optimization criteria. 
\\
\\
As our first example, we consider the approximation of a doubly-curved target surface using a generalized Miura-ori tessellation. Given a smooth target surface that we want to approximate, we consider two bounding surfaces displaced in the normal direction from the target surface (an upper and a lower bound) and construct a simple singly-corrugated strip in their interstice with one side of the strip lying on the upper surface and one side on the lower surface (see SI Section~S4A for more details). Then, applying our additive algorithm, we add strips to either side of the seed (and continuing on the growing patch) that approximately reflect the origami surface back and forth between the upper and lower target bounds, inducing an additional corrugation in a transverse direction to that of the corrugation in the seed. Fig.~\ref{fig:surface}A shows a high-resolution generalized Miura-ori sandwich structure of constant thickness obtained by our approach that approximates a mixed-curvature landscape, which would be very difficult to obtain using current techniques. As our second example, using a different DOF selection setup with no reference target surface, we grow a conical seed with a series of straight ridges with four-fold symmetry via facets that are created by reflections back and forth between a pair of rotating planes, shown in Fig.~\ref{fig:surface}B. As our third example, we turn to designing surfaces that have curve folds, folds that approximate a smooth spatial curve with non-zero curvature and torsion~\cite{huffman76primer}. Fig.~\ref{fig:surface}C shows a twisted version of David Huffman's \emph{Concentric Circular Tower}~\cite{figuring2004} obtained by our method, which uses a similar DOF selection setup as Fig.~\ref{fig:surface}B but begins with a high-resolution cone segment as the inner ring seed and follows with progressively thicker tilted cone rings added transversely (see SI Section~S4B for more details on both of these models). Fig.~\ref{fig:surface}D shows another curved fold model that uses the corollary to create a seed from a corrugated discrete planar parabola and proceeds to add new strips with constant flap angles and edge lengths, growing in a direction along the folds (see SI Section~S4C for more details). As our last surface example, we use our approach to create a disordered, crumpled surface that is isometric to the plane, again a structure that would be very difficult to obtain using current techniques. For each step of strip construction in the additive algorithm, the flap angle and edge lengths can be chosen randomly, thereby leading to a crumpled sheet that does not follow a prescribed MV pattern (Fig.~\ref{fig:surface}E). To model the physically realizable crumpled geometry, we have chosen flap angles according to the self-intersection bounds given by special vertex solutions along the entire front so that the growth of the sheet is a locally self-avoiding walk (see SI Section~S4D for more details).
\\
\\
Finally, to emphasize the flexibility of the corollary, we construct a quad strip that forms a folding connection between a canonically rough structure, a random walk in 3D, and a canonically smooth structure, a circle in 2D. Fig.~\ref{fig:surface}F shows a single folded strip generated by sampling a discrete Brownian path in 3D to form one boundary of a folded strip, choosing $k$ values such that its development falls on a circle and forms a single closed loop for any choice of flap angle value and choosing edge lengths such that the other boundary of the strip develops to another smaller concentric circle (see SI Section S4E). Indeed, the corollary allows for the freeform design of both folded ribbons and their pattern counterparts, independently. See SI Movies~S4--S9 for 3D animations of the models in Fig.~\ref{fig:surface}, and Figs.~S11, S12, S14, S16 and S17 for a gallery of other surface fitting, curved fold, disordered and Brownian ribbon results obtained by our additive approach.
\\
\\
Since the developability condition in Eq.~\eqref{eq:angle_sum} is always satisfied in our marching algorithm, all physical models created by it which do not self-intersect can transform from a 2D flat state to the isometric 3D folded state, typically through an energy landscape that includes geometric frustration (see SI Section~S5, Fig.~S18 and Movie~S10 for folding simulations). Because the landscape depends on the geometry of the folding pattern for which folding motions are not unique, this opens future routes to also program meta-stability. We also note that by replacing the right side in Eq.~\eqref{eq:angle_sum} by $K \ne 2 \pi$ and suitably modifying the subsequent trigonometric formulas, our additive approach generalizes to the design of non-Euclidean origami~\cite{berry2020topological,waitukaitis2020non}, with solutions uniquely existing in the same way when $K - \theta_{i,3} - \theta_{i,4} \in (\beta_i, 2\pi-\beta_i)$.
\\
\\
Overall, our study provides a unified framework for the inverse design of generic developable quad origami patterns and discrete developable surfaces via growth. A simple theorem forms the basis for a marching algorithm that replaces the solution of a difficult global optimization problem or case-specific geometric constructions with a scalable, easy-to-implement marching scheme for the evolution of a constrained folded front. This interplay between bulk rigidity and boundary flexibility that allows us to rapidly prototype computational designs of ordered, disordered, straight and curved folded geometries holds substantial promise for advances in discrete geometry, engineering applications and artistic creations alike.

% Bibliography
\bibliographystyle{Science_with_title}
\bibliography{additive_origami_bib}

\end{document}